\documentclass[12pt]{iopart}
\usepackage{iopams}  

\usepackage{amsmath,amsfonts,graphics,graphicx,epsfig,times,bbm}
\usepackage{amsfonts,graphics,times,bbm,amssymb}
\usepackage{epsfig}
\usepackage{graphicx}
\usepackage{color}
\usepackage{latexsym}
\usepackage{amscd}
\usepackage[all]{xy}

\newcommand{\odd}{\textup{Odd}} 
\newcommand{\cplane}[2]{(#1,#2)}
\def\qed{$\Box$}

\def\ens#1{\{#1\}}
\def\ie{\textit{i.e.}}
\def\eg{\textit{e.g.}}

\def\dag{^\dagger}

\def\mathvec#1{\mathbf{#1}}
\def\al{\alpha}
\def\ba{\beta} 

\def\rar{\rightarrow}

\def\lrar{\longrightarrow}

\def\slar#1{\stackrel{#1}{\lrar}}

\def\mcl{\mathcal} 
\def\mbb{\mathbb} 
 
\def\mfr{\mathfrak}
\newtheorem{prop}{Proposition}\def\PRO{\begin{prop}}\def\ORP{\end{prop}}
\newtheorem{coro}{Corollary}\def\COR{\begin{coro}}\def\ROC{\end{coro}}
\newtheorem{theo}{Theorem}\def\TH{\begin{theo}}\def\HT{\end{theo}}
\newtheorem{defi}[prop]{Definition}\def\DE{\begin{defi}}\def\ED{\end{defi}}
\newtheorem{lemme}[prop]{Lemma}\def\LE{\begin{lemme}}\def\EL{\end{lemme}}
\newcommand{\AR}[2][c]{$$\begin{array}[#1]{lllllllllllllll}#2\end{array}$$}

\def\EQ#1{\begin{eqnarray}#1\end{eqnarray}}
\def\hil#1{\mfr H_{#1}}

\def\ket#1{{|}#1\rangle}
\def\bra#1{\langle#1{|}}
\def\ztwo{{\mbb Z}_2}

\def\pit{\frac\pi2}

\def\cz#1#2{Z_{#1}^{#2}}
\def\cx#1#2{X_{#1}^{#2}}

\def\cx#1#2{X_{#1}^{#2}}
\def\phZ#1#2{N_{#1}}

\def\et#1#2{E_{#1#2}}

\def\GM#1#2#3{{M}_{#3}^{#1,#2}}

\def\al{\alpha}
\def\ba{\beta} 
\def\ga{\gamma}

\def\ta{\theta}

\begin{document}

\title[Generalized Flow and Determinism]{Generalized Flow and Determinism in  Measurement-based Quantum Computation}

\author{Daniel E. Browne}
\address{Departments of Materials and Physics, University of Oxford}
\ead{daniel.browne@merton.ox.ac.uk}
\author{Elham Kashefi}
\address{Christ Church College, University of Oxford}
\ead{elham.kashefi@comlab.ox.ac.uk}
\author{Mehdi Mhalla}
\address{Leibniz Laboratory, CNRS, Universit\'e de Grenoble }
\ead{mehdi.mhalla@imag.fr}
\author{Simon Perdrix}
\address{PPS, CNRS - Universit\'e Paris 7}
\ead{simon.perdrix@pps.jussieu.fr}

\begin{abstract}
We extend the notion of quantum information flow defined by Danos and Kashefi \cite{Flow06} for the one-way model \cite{RB01} and present a necessary and sufficient condition for the deterministic computation in this model. The generalized flow also applied in the extended model with measurements in the $\cplane X Y $, $\cplane X Z$ and $\cplane Y Z$ planes. We apply both measurement calculus and the stabiliser formalism to derive our main theorem which for the first time gives a full characterization of the deterministic computation in the one-way model. We present several examples to show how our result improves over the traditional notion of flow, such as geometries (entanglement graph with input and output) with no flow but having generalized flow and we discuss how they lead to an optimal implementation of the unitaries. More importantly one can also obtain a better quantum computation depth with the generalized flow rather than with flow. We believe our characterization result is particularly essential for the study of the algorithms and complexity in the one-way model. 
\end{abstract}


\maketitle

\section{Introduction}

Measurement-based quantum computation, and, more specifically, the one-way quantum computation model \cite{RB01} provides both a new theoretical description and a novel prescription for implementations of quantum computing. In the standard model of quantum computation -- the circuit model -- a computation is described as a network of unitary single and two-qubit gates acting on a register of qubits, followed, at the end, by measurement of each qubit. In the one-way quantum computation \cite{RB01}, first a special entangled multi-qubit state called a ``graph state''  or ``cluster state'' \cite{HDER06} is prepared, then the qubits are measured in a specified order and in specified bases.

An important aspect of this model is the way the inherent randomness of the measurement outcomes can be accounted for, so that the overall computation remains deterministic. This is accomplished by conditioning the basis of certain measurements upon the outcome of others, introducing a measurement order.

In \cite{RB01,RBB03} a prescription is given for the construction of deterministic measurement-patterns equivalent to any gate network. Nevertheless, one of the potential advantages of this new model is the possibilities it opens up for the development and design of quantum algorithms in a completely new picture, without direct recourse to the circuit model.

Thus an important question is, given a particular graph state and set of measurements, can these measurements be adapted in such a way that determinism of the computation is guaranteed? What is the structure of graph states and measurements which can be considered, such that the computation remains deterministic? In this paper, we provide a general framework to answer such questions.

Previously, a geometric condition on a graph state known as ``flow'', was developed which guaranteed that graph states satisfying a set of ``flow conditions'' would admit a deterministic computation, provided measurements were restricted to the $(X,Y)$ plane of the Bloch sphere \cite{Flow06}. Nevertheless, this condition was not necessary and did not take into account possibilities of measurements in other planes of the Bloch sphere, and the special way graph states transform under measurement of Pauli-operator observables \cite{graphstates}. In this article we provide a ``generalized flow'' condition. This is a set of geometric conditions on a graph, which are necessary and sufficient for that state to admit a deterministic one-way quantum computation under measurements in the $(X,Y)$, $(X,Z)$ and $(Y,Z)$ planes.

The structure of this article is as follows. We begin by reviewing the ``measurement calculus'' \cite{Mcal06} -- a means of algebraically representing measurement patterns in one-way quantum computation. After reviewing the flow condition introduced in \cite{Flow06}, we shall then present definitions of generalised flow and prove its properties. We show how this can be further generalised in the cases where certain qubits are restricted to Pauli measurements alone before concluding with some examples of the application of these concepts.

\section{Preliminary}

We briefly recall the definition of measurement patterns and various notions of determinism. More detailed introductions can be found in \cite{Nielsen05,Jozsa05,BB06}. In this paper, we will employ an algebraic approach towards measurement-based quantum computing (MQC) called, the \emph{Measurement Calculus} \cite{Mcal06} and its extension \cite{Perdrix06}. We define an arbitrary pure single qubit state by 
\AR{
\ket{+_{\ta,\phi}}=\cos(\frac{\ta}{2})\ket 0 + e^{i\phi}\sin(\frac{\ta}{2})\ket 1 \, ,
} 
and denote its orthogonal state (the opposite point in the Bloch Sphere) with 
\AR{
\ket{-_{\ta,\phi}}=\sin(\frac{\ta}{2})\ket 0 - e^{i\phi}\cos(\frac{\ta}{2})\ket 1\,,
} 
where $0\leq \ta \leq \pi$ and $0\leq \phi \leq 2\pi$. A computation, in the extended one-way model, is a combination of the following commands:
\begin{itemize}
\item 1-qubit preparations, $N_i$, to prepare the qubit $i$ in state $\ket{+}_i = \ket {+_{\pit,0}}_{i} $\,,
\item 2-qubit ``controlled $Z$'' entangling operators, $\et ij:=\Lambda Z_{ij}$\,,
\item 1-qubit destructive measurements, $\GM \lambda \alpha i$, on plane $\lambda\in \{\cplane XY,\cplane XZ, \cplane YZ\}$, defined by orthogonal projections into:
\begin{itemize}
\item $\ket {\pm_{\cplane XY,\alpha}}:=\ket{\pm_{\frac \pi 2,\alpha}}$ if $\lambda = \cplane XY$,
\item $\ket {\pm_{\cplane XZ,\alpha}}:=\ket{\pm_{\alpha,0}}$ if $\lambda = \cplane XZ$,
\item $\ket {\pm_{\cplane YZ,\alpha}}:=\ket{\pm_{\alpha,\frac \pi 2}}$ if $\lambda = \cplane YZ$,
\end{itemize}
 with the convention that $\ket{+_{\ta,\phi}}\bra{+_{\ta,\phi}}_i$ corresponds to the outcome $0$, while $\ket{-_{\ta,\phi}}\bra{-_{\ta,\phi}}_i$ corresponds to $1$\,,
\item 1-qubit Pauli corrections: $X_i$ and $Z_i$\,,
\end{itemize}
where $i$, $j$ represent the qubits on which each of these operations apply, and $0\leq \alpha \leq \pi$. Qubits are measured at most once, therefore we may represent unambiguously the outcome of the measurement outcome for qubit $j$ by $s_j$. To control the non-determinism of the measurement outcomes, certain corrections will depend upon previous measurement outcomes. These dependent corrections will be written as $U_i^{s_j}$, with $U_i^0=I$, and $U_i^1=U_i$. We will employ some set-theoretic notations, for example, $A^c$ denotes the complement of a subset $A$.

A \emph{measurement pattern} $\mathfrak P = (V,I,O,\lambda,S)$, or simply a pattern, is defined by the choice of $V$ a finite set of qubits, two possibly overlapping subsets $I\subseteq V$ and $O\subseteq V$ determining the pattern inputs and outputs, a finite sequence $S$ of commands acting on $V$, and a function  $\lambda: O^{c}\to \{\cplane XY,\cplane XZ, \cplane YZ\}$ which specifies the ``plane'' of the measurement on each measured qubits (i.e. the complement of the output qubits $O$). 
We will consider only \emph{runnable} patterns where no command depends on an outcome not yet measured, no command acts on a qubit already measured or not yet prepared (except preparation commands), and a qubit $i$ is measured (prepared) if and only if $i$ is not an output (input).

We are sometime interested to \emph{standardize} patterns: to put the commands sequence of the pattern in a particular order without changing the meaning of the computation. We choose an order where all the preparation commands appear first (i.e. right-most), then all the entanglement commands, followed by the measurements and then corrections.  The standardisation procedure is the basis for the new physical architecture proposed by measurement-based quantum computing where on performs all the entanglement in the beginning followed by local operation and classical communications. Furthermore, the rewriting of a pattern to standard form  allows one to check that a given pattern is runnable and it reveals parallelism in the pattern computation \cite{BK06}. 

We use commutation rules to interchange the order of operations to bring them into this form. Since the entanglement operations are to be performed first, when their order is interchanged with the correction operators they should not pick up the measurement-dependancy in these corrections. This can be guaranteed by ensuring that the entanglement command is in the normalizer group of the group generated by the correction groups. In order to commute the corrections to the end of the pattern we simply use the following equations:
$$\begin{array}{rcl}
M^{\cplane XY,\alpha}_i X_{i} &=&M^{\cplane XY,-\alpha}_i \\
M^{\cplane XY,\alpha}_i Z_{i} &=&M^{\cplane XY,\alpha+\pi}_i \\
M^{\cplane XZ,\alpha}_i X_{i} &=&M^{\cplane XZ,-\alpha+\pi}_i \\
M^{\cplane XZ,\alpha}_i Z_{i} &=&M^{\cplane XZ,-\alpha}_i \\
M^{\cplane YZ,\alpha}_i X_{i} &=&M^{\cplane YZ,\alpha+\pi}_i \\
M^{\cplane YZ,\alpha}_i Z_{i} &=&M^{\cplane YZ,-\alpha+\pi}_i \\

\end{array}$$

Therefore we have the following simple observation.
\PRO
Any one-way MQC model admits a standardization procedure if and only if the entanglement operator is normalizer of all the correction operators.
\ORP

We write $\hil I$ ($\hil O$) for the Hilbert space spanned by the inputs (outputs). The \emph{run}	 of a pattern consists simply in executing each command in sequence. If $n$ is the number of measurements (which is also the number of ``non-output'' qubits) then the run may follow $2^n$ different branches. Each branch is associated with a unique binary string  $\mathvec s$ of length $n$, representing the classical outcomes of the measurements along that branch, and a unique \emph{branch map} $A_{\mathvec s}$ representing the linear transformation from $\hil I$ to $\hil O$ along that branch. 

Branch maps decompose as $A_{\mathvec s}=C_{\mathvec s}\Pi_{\mathvec s}U$, where $C_{\mathvec s}$ is a unitary map over $\hil O$ collecting all corrections on outputs, $\Pi_{\mathvec s}$ is a projection from $\hil V$ to $\hil O$ (where $\hil V$ is the  Hilbert space spanned by all the qubits) representing the particular measurements performed along the branch, and $U$ is a unitary embedding (or isometry) from $\hil I$ to $\hil V$ collecting the branch preparations, and entanglements. Therefore
\AR{ 
\sum_{\mathvec s} A_{\mathvec s}\dag A_{\mathvec s}=\sum_{\mathvec s} U\dag\Pi_{\mathvec s}U=I
}
and $T(\rho):= \sum_{\mathvec s} A_{\mathvec s}\rho A_{\mathvec s}\dag$ is a trace-preserving completely-positive map (cptp-map), explicitly given as a Kraus decomposition. One says that the pattern \emph{realizes} $T$. 
 
A pattern is said to be \emph{{deterministic}} if it realizes a cptp-map that sends pure states to pure states. This is equivalent to saying that branch maps are proportional.
A pattern is said to be \emph{strongly deterministic} when branch maps are equal (up to a global phase), \ie, for all $\mathvec s_1$, $\mathvec s_2\in \ztwo^n$, $A_{\mathvec s_1}=e^{i \phi_{s_1,s_2}}A_{\mathvec s_2}$. 
A pattern is said to be \emph{uniformly deterministic} if it is deterministic for all values of its measurement angles. Finally a pattern is said to be \emph{stepwise deterministic} if it is deterministic after performing each single measurement together with all the corrections depending on the result of that measurement. 


The main result of our paper is a necessary and sufficient condition for strong uniform determinism based on the geometry of the entanglement structure which underlies a measurement pattern. Let us define an \emph{open graph state} $(G,I,O, \lambda )$ to be a state associated with an undirected graph $G$ together with two subsets of nodes $I$ and $O$, called inputs and outputs. We write $V$ for the set of nodes in $G$, $I^c$, and $O^c$ for the complements of $I$ and $O$ in $V$, $N_{G}(i)$ for the set of neighbours of $i$ in $G$, $i\sim j$ for $(i,j) \in G$, and $E_G:=\prod_{i \sim j}E_{ij}$ for the global entanglement operator associated to $G$. We first recall the definition of \emph{flow}, under which one can construct a set of dependent corrections such that the obtained pattern is strongly and uniformly deterministic \cite{Flow06}. 

\DE An open graph state $(G,I,O,\lambda)$, such that $\forall i\in O^c, \lambda(i)=\cplane XY$, has \emph{flow} if there exists a map $f:O^c\rar I^c$ (from measured qubits to prepared qubits) and a partial order $>$ over $V$ such that for all $i\in O^c$:
\\--- (F1) $i \sim f(i)$\,,
\\--- (F2) $i<f(i)$\,,
\\--- (F3) $\forall k\in N_{G}(f(i))\setminus\ens i$\; we have\; $i<k$\,.
\ED

As one can see, a flow consists of two structures: a function $f$ over vertices and a matching partial order over vertices. In order to obtain a deterministic pattern for an open graph state with flow, dependent corrections will be defined based on function $f$. The order of the execution of the commands is given by the partial order induced by the flow. The matching properties between the function $f$ and the partial order $>$ will make the obtained pattern runnable.

\TH \label{t-flow} \cite{Flow06} Suppose the open graph state $(G,I,O,\lambda)$, such that  $\forall i\in O^c, \lambda(i)=\cplane XY$,  has flow $(f,>)$, then the pattern:
\AR{
\mfr P_{f,G}&:=&\prod^>_{i\in O^c}\left(\cx{f(i)}{s_i}\cz{N_{G}(f(i))\setminus\ens{i}}{s_i}\;\GM {\cplane X Y} {\alpha_i} i\right)\;E_G \;N_{I^c} \,,
}
where the product follows the dependency order $>$, is runnable, uniformly and strongly deterministic, and realizes the unitary embedding:
\AR
{U_{G}&:=&
\left(\prod_{i\in O^c}\bra{{+_{\cplane XY,\alpha_i}}}_i\right)\, E_G\phZ {I^c}0\,.
}
\HT
The above theorem provides a necessary condition 
 for determinism for the one-way model considering only measurements in the $\cplane XY$ plane, which encompasses for example, the measurement patterns proposed \cite{RB01,mqqcs}. Nevertheless, it can be useful to construct patterns which contain measurements in other planes \cite{BB06}, and this arises naturally when one uses the graph transformation rules associated with Pauli measurements \cite{graphstates} to reduce the size of a pattern. As we shall describe in this article one can extend the notation of flow to obtain a necessary and sufficient condition considering measurements in all the $\cplane XY$, $\cplane XZ$ and $\cplane YZ$ planes. This will lead to a full characterization of deterministic computation in the MQC models. As a result we also obtain a tight bound on depth complexity that improves the presented results in \cite{BK06}.


\section{Generalized Flow}

In order to describe the motivation behind our construction of the generalized flow we first briefly explain the main idea behind the proof of the flow theorem (Theorem \ref{t-flow}). Recall that the \emph{graph stabiliser}~\cite{graphstates,NC00} at qubit $i$ is defined as $K_i:=X_{i}(\prod_{j\in G(i)}Z_j)$ and one has the following relation:
\EQ{
\label{e-stab}
K_i E_G \phZ {I^c}0=E_G \phZ {I^c}0\,.
}

The proof of Theorem \ref{t-flow} is based on the following simple observation. We could make a measurement $\GM {\cplane XY} \alpha i$ ``deterministic'' (corrected) if it could be pre-composed by  an anachronical $Z_i^{s_i}$ correction (i.e. conditioned on the outcome of a measurement which hasn't happened yet). This  unphysical scenario is a useful starting point for our proof.
\AR{
\bra{{+_{\cplane XY, \alpha}}}_i&=&\GM {\cplane XY} \alpha i \cz i {s_i}\,.}
The flow construction guarantees that such a deterministic pattern with anachronical corrections can be transformed into a runnable pattern, where all dependencies now do respect the proper causal ordering.
 It is easy to verify that, the pattern $\mfr P_{f,G}$ in Theorem \ref{t-flow} can be equivalently written in terms of  anachronical measurements as:
\AR{
\mfr P_{f,G}&=&\prod^>_{i\in O^c}\left(\GM {\cplane XY} {\alpha_i}i \cz i{s_i}K_{f(i)}^{s_i}\right)E_G \phZ {I^c}0\,.
}
The key observation which allows us to tranform this into a runnable pattern is that the flow conditions mean that there exists a stabiliser $K_{f(i)}$ which when composed with the anachronical correction, forms an operator which commutes with the measurement, and thus the pattern can be brought into runnable order.

A natural way to extend this idea is to consider a set of vertices  as a \emph{correcting set}. Hence instead of working with a function $f: O^c \rightarrow I^c$ defining the correcting vertices, we will have a function $g: O^c \rightarrow \mcl{P}^{I^c}$ defining the correcting sets of vertices, where $\mcl{P}^{I^c}$ denotes the power set of all the subsets of vertices in $I^c$. It is important to note that the condition on these correcting sets will depend on the plane which the measurement will be performed, as measurement in different planes require a different anachronical correction. We define the odd neighborhood of a set of vertices $K$ to be the set $\odd (K)=  \{u\, ,\,  |N_G(u)\cap K|=1 \mod 2\}$.  

\begin{figure}
\begin{center}
\includegraphics[scale=0.5]{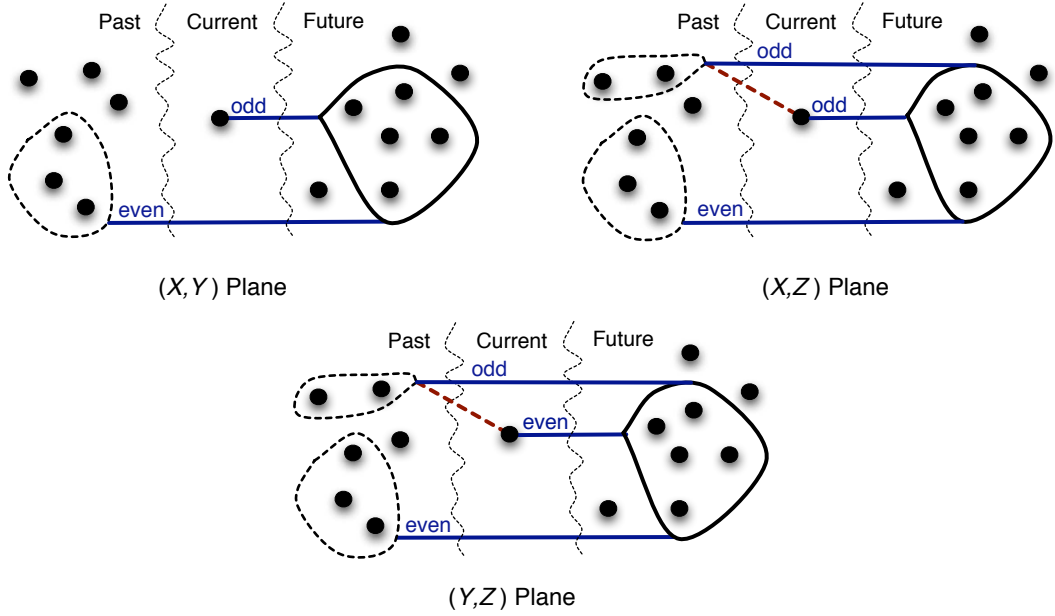}
\caption{The pictorial presentation of the generalised flow conditions (G1-G5) for different measurement planes. The straight lines (blue) stand for multiple edges in the entanglement graph where the labels give the parity of the number of these connections and the doted straight lines (red) for a single edge. The single qubit in Current layer denote the qubit to be measured, its correcting set lays in Future layer (black closed curve). The neighbours of the correcting set belonging to Past layer are denoted by doted closed cure. }
\label{f-correctingset}
\end{center}
\end{figure}

\DE \label{d-gflow} An open graph state $(G,I,O,\lambda)$ has \emph{generalized flow} if there exists a map $g: O^c \rightarrow \mcl{P}^{I^c}$ (from measured qubits to a subset of prepared qubits) and a partial order $<$ over $V$ such that for all $i\in O^c$, 
\\---(G1)\ if $j\in g(i)$ and $i\neq j$ then $i<j$,
\\---(G2)\ if $j\le i$ and $i\neq j$ then $j\notin \odd(g(i))$,
\\---(G3)\ if $\lambda(i)=\cplane XY$ then $i\notin g(i)$ and $i\in \odd(g(i))$,
\\---(G4)\ if $\lambda(i)=\cplane XZ$ then $i\in g(i)$ and $i\in \odd(g(i))$,
\\---(G5)\ if $\lambda(i)=\cplane YZ$ then $i\in g(i)$ and $i\notin \odd(g(i))$.
\ED
One can consider the partial order $<$ as a notion of time. Then, condition (G2)  says that all the vertices with an odd number of connection to the correcting set $g(i)$ should belong to the past of $i$. These conditions can be better understood pictorially, as illustrated in  Figure \ref{f-correctingset}. Similar to the Theorem \ref{t-flow} we will apply dependent stabiliser corrections on all the qubits in the correcting set. The evenness or oddness condition on the number of the connections between a vertex and its correcting set and  neighbors will guarantee that the  anachronical correction on qubit $i$ can be transformed to a correction with casual dependencies.

\section{Determinism Theorem}

A necessary and sufficient condition for determinism in the extended one-way model is given in the following two theorems. It is important to note that this condition can be easily extended to any other MQC models (\eg~teleportation-based models \cite{Gott99,L03}), since there exist compositional embeddings from the one-way model to all other MQC models \cite{Mcal06}. Recall that $g(i)$, where $g$ is a generalised flow, is a subset of vertices.
\TH\label{t-forward}
Suppose the open graph state $(G,I,O, \lambda)$ has generalised flow $(g,>)$, then the pattern:
\AR{
\mfr P_{g,G}&:=&\prod^>_{i\in O^c}\left(\cx{g(i)\setminus \{i\}}{s_i}\cz{\odd(g(i))}{s_i}\;\GM {\lambda(i)} {\alpha_i} i\right)\;E_G N_{I^c} \,,
}
where the product follows the dependency order $>$,  is runnable, uniformly, strongly and stepwise deterministic, and realizes the unitary embedding:
\AR
{U_{G}&:=&
\left(\prod_{i\in O^c}\bra{{+_{\lambda(i),\alpha_i}}}_i\right)\, E_G\phZ {I^c}0\,. 
}
\HT

\TH\label{t-backward}
Suppose the pattern $\mfr P$ is uniformly, strongly  and stepwise deterministic, then the underlying geometry of $\mfr P$ has generalized flow and the pattern realizes the unitary embedding:
\AR
{U_{G}&:=&
\left(\prod_{i\in O^c}\bra{{+_{\lambda(i),\alpha_i}}}_i\right)\, E_G\phZ {I^c}0\,. 
}

\HT

The next lemma will be used in the proof of Theorem \ref{t-backward} and illustrates the role that the strong condition of uniformity will play. Denote by $P_{\ket \psi}$ a projection to state $\ket \psi$.

\LE \label{l-uniform}
If for all $\alpha$ in the  $(X,Z)$,$(X,Y)$ or $(Y,Z)$ plane $P_\alpha \ket{\psi}= e^{if(\alpha)} P_\alpha \ket{\psi'}$ then $\ket{\psi}=e^{i \theta} \ket{\psi'}$
\EL
{\bf Proof.} We write the proof for the case of a projection in $(X,Z)$ plane as other cases are similar. It suffices to consider the angles of $\alpha=\{0,\pi/2, \pi\}$, or in other words, measurements of $X$ and $Z$ observables. First we write the states in the basis of the eigenvectors of $Z$: 
\AR{
\ket{\psi}= a \ket{0} \ket{\psi_0} + a' \ket{1} \ket{\psi_1} \;\;\; ,\;\;\;  \ket{\psi'}= b \ket{0} \ket{\psi'_0} + b' \ket{1} \ket{\psi'_1}\, .
}
The lemma condition with $\alpha=0$ and $\alpha=\pi$ (i.e. projections onto $\ket{0}$ and $\ket{1}$) implies that
\AR{
\ket{\psi_0} =\ket{\psi'_0} \;\;\; , \;\;\; \ket{\psi_1} =\ket{\psi'_1} \, .
} 
We also obtain $a=e^{i\phi_0}b$ and  $a'=e^{i\phi_1}b'$ and therefore 
\AR{
\ket{\psi'}= e^{i\phi_0}(a \ket{0} \ket{\psi_0} + e^{i (\phi_1- \phi_0)}a' \ket{1} \ket{\psi_1})\, .
}
Now consider the projection  onto $\ket +$ state ($\alpha=\pi/2$) which implies
\AR{
a \ket{\psi_0}+ a' \ket{\psi_1}= \e^{i \theta} (a \ket{\psi_0}+ a' e^{i (\phi_1- \phi_0)} \ket{\psi_1}) \, .
} 
Thus $e^{i (\phi_1- \phi_0)}=1$ or $a'=0$ which completes the proof. \qed

The following equations on anachronical corrections play the central rule in the proof of theorem \ref{t-forward}:
\EQ{\label{e-XY}
\bra{{+_{\cplane XY, \alpha}}}_i&=&\GM {\cplane XY} \alpha i \cz i {s_i}\, ,\\
\label{e-XZ}
\bra{{+_{\cplane XZ, \alpha}}}_i&=&\GM {\cplane XZ} \alpha i \cx i {s_i}\cz i {s_i}\, ,\\
\label{e-YZ}
\bra{{+_{\cplane YZ, \alpha}}}_i&=&\GM {\cplane YZ} \alpha i \cx i {s_i}\, ,
}
 
\noindent {\bf Proof of Theorem \ref{t-forward}.} We prove one case, where all the measurements are assumed to be in $\cplane YZ$ plane, all other cases have a similar proof. Clearly, the generalized flow conditions make the following pattern a runnable one,
\AR{
\mfr P_{g,G}:=\prod^>_{i\in O^c}\left(\cx{g(i)\setminus \{i\}}{s_i}\cz{\odd(g(i))}{s_i}\;\GM {\cplane YZ} {\alpha_{i}} i\right)\;E_G \phZ {I^c}0 \, .
}
By commuting the corrections and using the definition of the graph stabiliser, $K_i$, we have:
\AR{
\mfr P_{g,G}=\prod^>_{i\in O^c}\GM {\cplane YZ} {\alpha_{i}} i \; X_i^{s_i} \; K_{g(i)}^{s_i} E_G \phZ {I^c}0\,.
}
Note that to derive the above equality we have also used the trivial equations $Z_j^{s_i}Z_j^{s_i}=I$ to complete the missing part of any stabiliser. Recall that Condition (G2) for $\cplane YZ$ measurement will guarantee to have the required even number of such missing $Z$ operators. Finally from equations \ref{e-stab} and \ref{e-YZ} we obtain the following uniformly, strongly and stepwise deterministic pattern 
\AR{
\mfr P_{g,G}=(\prod_{i\in O^c}\bra{{+_{(Y,Z),\alpha_{i}}}}_i)\, E_G\phZ {I^c}0\,. 
}\qed

\noindent{\bf Proof of Theorem \ref{t-backward}.} 
We start from the end of the pattern computation (\ie~last measurement commands). Let $P_{O^C}$ be the projector over the state $ \Pi_{i \in O^C} \ket{+_{\lambda(i),\alpha_i}}$. Suppose the last measurement is in the plane $\cplane XY$ and is performed on qubit $n$, it creates then the following two branches 
\AR{
\ket \Psi
&\slar{M^{\cplane XY,\alpha_n}_n}&
\left\{
\begin{array}{l}
P_{O^C} E_G N_{V\setminus I}  \ket{\psi}_I \;\;\;\;\;\;\;\;\;\; s_n=0\\\\
P_{O^C} Z_n E_G N_{V\setminus I}  \ket{\psi}_I \;\;\;\;\;\; s_n=1
\end{array}
\right.
}
where we have used the fact $\bra {-_{\al_n}}Z_n=\bra {+_{\al_n}}$.  Now from the stepwise determinism there exists a collection of correction $C_A$ on output qubit such that 
\AR{
C_A P_{O^C} Z_{n} E_G N_{V\setminus I}  \ket{\psi}_I = P_{O^C} E_G N_{V\setminus I} \ket{\psi}_I \, ,
} 
since the corrections are performed on output qubit we can commute $C_A$ with $P_{O^C}$ and write
\AR{
P_{O^C} C_{A} Z_{n} E_G N_{V\setminus I}  \ket{\psi}_I = P_{O^C} E_G N_{V\setminus I} \ket{\psi}_I \, .
}
The above equation is valid for any value of $\al_n$ (uniformity condition) and thus according to lemma \ref{l-uniform},  $C_{A} Z_{n} E_G N_{V\setminus I}  \ket{\psi}_I = E_G N_{V\setminus I} \ket{\psi}_I$, so $C_{A}Z_{n}$ stabilises the state  $E_G N_{V\setminus I} \ket{\psi}_I$ and since it is a Pauli operator it can be written as a product of the Pauli group generators (graph stabiliser, $K_i$), thus there exists a set $S\subseteq V$ such that 
\AR{
C_{A}Z_{n}=\Pi_{u\in S} X_u Z_{N(u)} \,.
} 
It remains to prove that $S$ is indeed the correcting set of qubit $n$ and satisfies the condition of the generalised flow conditions. First we show that $S\cap I = \emptyset$, let $\ket G = E_G N_{V\setminus I} \ket{+}_I$, then for any arbitrary subset $K\subseteq I$ since $A\cap I=\emptyset$ we have
\AR{
C_A Z_n E_G N_{V\setminus I}  Z_K \ket{+}_I = C_A Z_n Z_K \ket{G}= Z_K \ket{G} \, .
}
In particular for $K=\emptyset$ we obtain $C_A Z_n \ket{G}= \ket{G}$. Suppose now there exists $e \in S\cap I $, and set $K=\{e\}$, since $e\in S$ then $Z_e$ anti commutes with $C_{A}$ and therefor
\AR{
C_{A}Z_{n} Z_{e}\ket G = -Z_{e}C_{A}Z_{n}\ket G = -Z_e\ket G \, ,
} 
which leads to a contradiction and proves $S\cap I = \emptyset $. 

On the other hand since $A$ does not intersect with the set of already measured qubits any $u \in S$ cannot act on a measured qubit as it will not be simplified later and $X_u$ on the other side will appear. Also $N(u)$ should see measured qubits evenly so that $Z_i$ on the previously measured qubit will cancel out each other as well. Therefore $S$ is the correcting set for the qubit $n$ in terms of Definition \ref{d-gflow}. The presented argument can be similarly carried out for all the measurements performed in the previous stages which completes the proof.\qed

\section{Pauli Measurements}

Pauli measurements play a central role in one-way quantum computing. In particular, it is known that the action of such a measurement on a graph state is to leave the remaining qubits in a graph state (up to a local Clifford-group correction) \cite{graphstates}. Definition \ref {d-gflow} provides conditions for determinism when single qubits at any angle in specified Bloch-sphere planes are allowed. The special properties of  Pauli measurements (for example, that they simultaneously lie in two measurement-planes) mean that if one restricts the measurement of certain qubits to certain specific Pauli measurements, one must extend the generalised flow conditions in order to account for these extra properties.

In this section, we introduce such an extension. We will use the convention that the labeling function $\lambda(i)$ for any non output qubit $i$, is either a plan -- $\cplane XY$, $\cplane XZ$, or $\cplane YZ$ -- or a vector -- $X$, $Y$, or $Z$ (\ie~Pauli measurements). First, notice that a Pauli measurement, say $X$, can be interpreted as a $\cplane XY$ or $\cplane XZ$ measurement and thus it may satisfies the conditions of either a $\cplane XY$ or a $\cplane XZ$ measurement. Second, when a qubit is measured according to a Pauli operator, say $X$, then, after the measurement, the state of this qubit takes $\pm X$ as its stabiliser. We use this property to allow already-measured qubits to be included in a correcting set. Finally the following relation between Pauli correction and Pauli measurements will be used for the Pauli flow construction
\EQ{
\label{e-XX}
M^X X&=&M^X\\
\label{e-YY}
M^Y Y&=&M^Y\\
\label{e-ZZ}
M^Z Z&=&M^Z
} 

\DE An open graph state $(G,I,O,\lambda)$ has \emph{Pauli flow} if there exists a map $p: O^c \rightarrow \mcl{P}^{I^c}$ (from measured qubits to a subset of prepared qubits) and a partial order $<$ over $V$ such that for all $i\in O^c$,
 \\---(P1)\ if $j\in p(i)$, $i\neq j$, and $\lambda(j)\notin \{X,Y\}$ then $i<j$,
\\---(P2)\ if $j\le i$, $i\neq j$, and $\lambda(j)\notin \{Y,Z\}$  then $j\notin \odd(p(i))$,
\\---(P3)\ if $j\le i$, $j\in p(i)$ and $\lambda(j)=Y$ then $j\in \odd(p(i))$,
\\---(P4)\ if $\lambda(i)=\cplane XY$ then $i\notin p(i)$ and $i\in \odd(p(i))$,
\\---(P5)\ if $\lambda(i)=\cplane XZ$ then $i\in p(i)$ and $i\in \odd(p(i))$,
\\---(P6)\ if $\lambda(i)=\cplane YZ$ then $i\in p(i)$ and $i\notin \odd(p(i))$,
\\---(P7)\ if $\lambda(i)=X$ then $i\in \odd(p(i))$,
\\---(P8)\ if $\lambda(i)=Z$ then $i\in p(i)$,
 \\---(P9)\ if $\lambda(i)=Y$ then either:\;\; $i\notin p(i) \; \& \; i\in \odd(p(i))$ \;\; or \;\; $i\in p(i) \; \& \; i\notin \odd(p(i))$.
 
\ED

\TH\label{t-forward2}

Suppose the open graph state $(G,I,O, \lambda)$ has Pauli flow $(g,>)$, then the pattern:
\AR{
\mfr P_{g,G}&:=&\prod^>_{i\in O^c}\left(\cx{g(i)\cap \{j\, , \, j> i\}}{s_i}\cz{\odd(g(i))\cap \{j\, , \, j>i\}}{s_i}\;\GM {\lambda(i)} {\alpha_i} i\right)\;E_G N_{I^c} \,,
}
where the product follows the dependency order $>$, is deterministic and realizes the unitary embedding:
\AR
{U_{G}&:=&
\left(\prod_{i\in O^c}\bra{{+_{\lambda(i),\alpha_i}}}_i\right)\, E_G\phZ {I^c}0\,. 
}
\HT

\textbf{Proof:} The proof is similar to the proof of theorem \ref{t-forward}. 
In $(P1)$, if $\lambda(j)\in \{X,Y\}$, $j$ may be in the $p(i)$ even if $j\le i$ since $M_{i}^X X_{i}=M_{i}^X$ and $M_{i}^Y X_{i}Z_{i}=M_{i}^Y$. Notice that if $\lambda(j)=Y$, $j\le i$ and $j\in p(i)$ then $j$ must be in $\odd(p(i))$ 
-- $(P3)$ -- because of the $Z_{i}$ command in $M_{i}^Y X_{i}Z_{i}=M_{i}^Y$.
In $(P2)$, if $\lambda(j)=Z$, then $j$ may be in $\odd(p(i))$ even if $j\le i$, since $M_{i}^ZZ_{i}=M_{i^Z}$. The condition $\lambda(j)\neq Y$ in $(P2)$ is necessary because of $(P3)$.
Finally, $(P7),(P8)$, and $(P9)$ are obtained from $(P4), (P5)$, and $(P6)$ since a $X$ measurement is both a $\cplane XY$ and a $\cplane XZ$ measurement, and so on.\qed

\section{Examples}

\begin{figure}
\begin{center}
\includegraphics[scale=0.35]{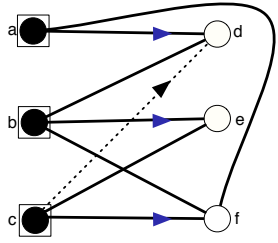}
\caption{A graph with generalised flow but no flow: $g(a)=d, g(b)=e, g(c)=\{d,f\}$. The blue arrows represent the flow edges, where as black arrow indicates a virtual flow edge, (an edge that is not an edge of the graph state).}
\label{f-noflow}
\end{center}
\end{figure}

Trivially, any open graph state with a flow also has a generalised flow, but the following set of examples show how the generalised flow can be beneficial. The open graph state in Figure \ref{f-noflow} has no flow (due the cyclic connections), but it admits a generalized flow. This example demonstrates the fact having flow is not a necessary condition for uniform determinism, contrary to the existence of the generalized flow, as it is shown in theorem \ref{t-backward}. 

\begin{figure}
\begin{center}
\includegraphics[scale=0.35]{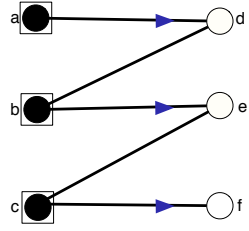}
\caption{A graph with flow: $f(a)=d, f(b)=e, f(c)=f$. The blue arrows represent the flow edges and the partial order over the vertices is $a<\{b,d\}<\{e,c\}<f$.}
\label{f-flowcircuit}
\end{center}
\end{figure}

\begin{figure}
\begin{center}
\includegraphics[scale=0.35]{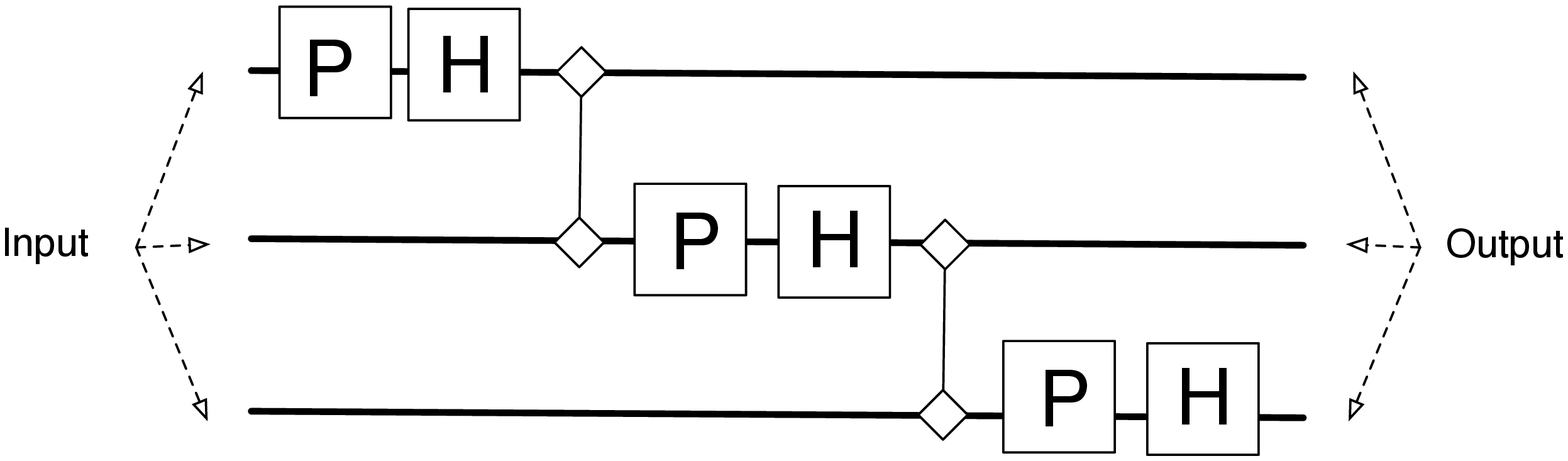}
\caption{The circuit implementation of the pattern in Figure \ref{f-flowcircuit}, with controlled-$Z$, phase $P(-\al)$ and Hadamard $H$ gates.}
\label{f-circuit}
\end{center}
\end{figure}

In relation to the circuit model, having a generalised flow lead to an optimal realisation. It is known that given a pattern where the underlying geometry has flow one can directly decompose the pattern into a circuit with no auxiliary qubits that implements the same unitary \cite{Flow06}. Consider for example the pattern given in Figure \ref{f-flowcircuit} that implements the following pattern (all measurement are in $(X,Y)$ plane)
\AR{
\cx {s_c} f \cx {s_c} d  M^\ga_c \;\; \cx {s_b} e \cz {s_b} c M^\ba_b \;\; \cx {s_a} d \cz {s_a} b M^\al_a \;\; \et ad \et db \et be \et ec \et cf \;\; N_d N_e N_f
}
that can be decomposed to the circuit given in Figure \ref{f-circuit} using the construction of \cite{Flow06}. The base of this procedure is to replace the pattern $\cx {s_i} j M^\al_i\et ij N_j$ with Phase and Hadamard gate and the remaining edges of the graph with ctrl-$Z$ gates.

\begin{figure}
\begin{center}
\includegraphics[scale=0.35]{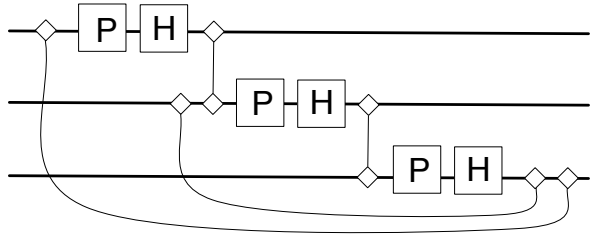}
\caption{The acausal circuit implementation of the pattern in Figure \ref{f-noflow}.}
\label{f-acausalcircuit}
\end{center}
\end{figure}

Now if we follow the same construction for pattern in Figure \ref{f-noflow} that has generalised flow but no flow, in order to remove all the auxiliary qubits, we obtain acausal circuit (not runnable), Figure \ref{f-acausalcircuit}. Of course, there exists another casual circuit implementing the same unitary but it might need more gates. This suggest that how patterns with generalised flow may implement more efficiently a given unitary. 

\begin{figure}
\begin{center}
\includegraphics[scale=0.35]{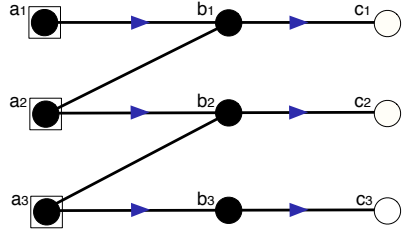}
\caption{Open graph state having a flow of depth $5$ and a generalized flow of depth $2$.}
\label{f-depthflow}
\end{center}
\end{figure}

Even if a graph has a flow, looking for a generalized flow is a way to decrease the depth of the computation, \ie~the parallel execution time. For instance, the open graph state  in Figure \ref{f-depthflow} has a flow function $f$ given below
\AR{
f(a_i)=b_i \;\;\; \&\;\;\; f(b_i)=c_i \, ,
} 
with the corresponding partial order being $a_1 < \{b_1, a_2\}<\{c_1,b_2,a_3\}<\{c_2,b_3\}< c_3$ and hence the depth of this flow is $5$. We know the flow function $f$ is unique \cite{Beaudrap06a} and since each $b_i$ must be greater than $a_{i}$ and each $a_{i}$ is greater than $b_{i-1}$, thus the minimal depth is $5$. However, this open graph state has also a generalized flow defined as 
\AR{
g(a_{1})=\{b_1,b_2\} \;\;\; \& \;\;\; g(a_{2})=\{b_2,b_3\} \;\;\; \& \;\;\; g(a_{3})=b_3 \;\;\; \& \;\;\; g(b_{i})=c_i \, ,
}
with partial order $\{a_1,a_2,a_3\} < \{b_1,b_2,b_3\}$ with depth $2$. On can easily extend this example (Figure \ref{f-depthflow}) to construct for any given $n$, an open graph state with $3n$ vertices having no flow of depth less than $n+1$, but a generalized flow of depth $2$.

\begin{figure}
\begin{center}
\includegraphics[scale=0.35]{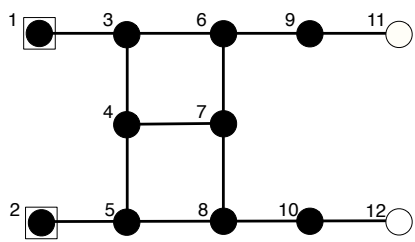}
\caption{Open graph state having no generalized flow but having a Pauli flow}
\label{f-pauliflow}
\end{center}
\end{figure}

Our final examples deal with the case of Pauli flow. An open graph with no generalized flow is not uniformly deterministic, but it can be still deterministic if one restrict some of the angel of measurements to Pauli. For instance, the open graph state in Figure \ref{f-pauliflow}, has no generalized flow, but it has the following Pauli flow with depth 1 when all the non output qubits are $X$-measured (implementing SWAP operator).
\AR{
p(1)&=&\{3,7,10\}\\
p(2)&=&\{5,7,9\}\\
p(3)&=&\{4,8,10\}\\
p(4)&=&\{7,9,10\}\\
p(5)&=&\{4,6,9\}\\
p(6)&=&\{9\}\\
p(7)&=&\{4,6,8,9,10\}\\
p(8)&=&\{10\}\\
p(9)&=&\{11\}\\
p(10)&=&\{12\}\\
}

\begin{figure}
\begin{center}
\includegraphics[scale=0.35]{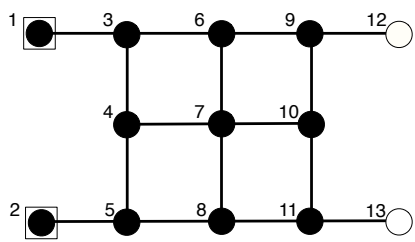}
\caption{An open graph state with no Pauli or generalised flow implementing a deterministic pattern for the SWAP operator.}
\label{f-nopauli}
\end{center}
\end{figure}

Finally, Pauli flow is not necessary for determinism in a graph-state pattern whose measurements are solely Pauli - since these do not require any adaptive measurements. An example of this is the open graph state in Figure \ref{f-nopauli}, has no Pauli flow but is deterministic and realizes also the SWAP operator \cite{RBB01SWAP}.

\section{Conclusion}

What makes the measurement-based quantum computing special is the fact that one can employ probabilistic measurement operators and yet perform a deterministic computation by imposing a causal dependent structure over the measurements sequence. On the other hand the MQC highlights the role of entanglement as a resource for quantum computing. Hence a full understanding of the MQC  depends on gaining insight into the interplay of these two ingredients.

In this article we have extended the notion of flow \cite{Flow06} on the geometry of the entanglement graph required for a one-way computing to address the above questions. We have presented for the first time a full characterisation of deterministic computation in the one-way model independent of any reference to the circuit model. Having generalised flow is a necessary and sufficient condition for  uniform determinism. On the other hand if one is willing to restrict to a particular set of angles such as Pauli measurements then the Pauli flow criteria might be used.

One interesting consequence of patterns with generalised flow (but no flow) is that they can admit very compact implementations of a given unitaries (as illustrated in our examples). In particular, the generalised flow admits a great deal of flexibility in the causal structure of the corrections which can have have little in common with the structure of the associated quantum circuit. Further investigation of such features will be a line of future research.

A further important open question is how one can design an efficient algorithm to find generalised flow given an open graph state, which would aid direct pattern design (as proposed in \cite{BDK06}). In the particular case where $|I|=|O|$, if a geometry has flow it is unique and can be found efficiently \cite{BDK06,Beaudrap06a,Beaudrap06b}, using a combination of network flow algorithms, and Tarjan's algorithm to avoid acausal sequences of measurements (forming ``vicious circuits''), as described in detail in \cite{Beaudrap06a,Beaudrap06b}.

We conjecture that an algorithm for generalised flow will be similar to  \cite{Beaudrap06a,Beaudrap06b}. First one would find a maximal collection of disjoint Input-Output paths to attempt to define the flow function. Then if the obtained partial order has vicious circuit one can attempt to cancel its effect  using the additional vertices in the correcting set.

We believe this work can form a basis for the development of novel quantum algorithms conceived solely in the language of measurement-based quantum computation.

\section{Acknowledgements}

E.K. would like to thank Vincent Danos and  Damian Markham	 for insightful discussions about flow and its possible extension and also for early proof of some of the presented results in this article.

This work was supported by Christ Church, Oxford, Merton College, Oxford, the EPSRC's QIPIRC programme and the EU-funded QICS network.


\end{document}